\documentclass{aa}                   
\usepackage{graphicx}                
\usepackage{txfonts}                 

\newcommand{\Halpha}{H$\alpha$}
\newcommand{\kms}{km\,s$^{-1}$}
\newcommand{\ms}{m\,s$^{-1}$}

\begin{document}

\title{The spectroscopic orbit of Capella revisited\thanks{Based on data obtained with the STELLA robotic
telescope in Tenerife, an AIP facility jointly operated by AIP and
IAC.}\fnmsep\thanks{Table \ref{T1} is only available in electronic form
at the CDS via anonymous ftp to cdsarc.u-strasbg.fr (130.79.128.5)
or via http://cdsweb.u-strasbg.fr/cgi-bin/qcat?J/A+A/}}

\author{M.~Weber \and K. G.~Strassmeier}

\offprints{K. G. Strassmeier}

\institute{Leibniz-Institut f\"ur Astrophysik Potsdam (AIP), An der Sternwarte
16, D-14482 Potsdam, Germany\\
\email{MWeber@aip.de},\email{KStrassmeier@aip.de}}

\date{Received ... ; accepted ...}

\abstract{Capella is among the few binary stars with two evolved
giant components. The hotter component is a chromospherically active
star within the Hertzsprung gap, while the cooler star is possibly
helium-core burning.}{The known inclination of the orbital plane
from astrometry in combination with precise radial velocities will
allow very accurate masses to be determined for the individual
Capella stars. This will constrain their evolutionary stage and
possibly  the role of the active star's magnetic field
on the dynamical evolution of the binary system.}{We obtained a
total of 438 high-resolution \'echelle spectra during the years 2007--2010 
and used the measured velocities to
recompute the orbital elements. Our double-lined orbital
solution yields average residuals of 64~\ms\ for the cool
component and 297~\ms\ for the more rapidly rotating hotter
component.}{The semi-amplitude of the cool component is smaller by
0.045~\kms\ than the orbit determination of Torres
et al. from data taken during 1996--1999 but more precise by a
factor of 5.5, while for the hotter component it is larger by
0.580~\kms\ and more precise by a factor of 3.6. This corresponds to masses
of 2.573$\pm$0.009~$M_\odot$ and 2.488$\pm$0.008~$M_\odot$ for the
cool and hot component, respectively. Their relative errors of
0.34\% and 0.30\% are about half of the values given in Torres et al.
for a combined literature-data solution but with absolute values
different by 4\% and 2\% for the two components, respectively.
The mass ratio of the
system is therefore $q=M_A/M_B=0.9673\pm0.0020$.}{Our orbit is the
most precise and also likely to be the most accurate ever obtained for
Capella.}



\maketitle

\section{Introduction}

The spectrum of the \object{Capella} ($\alpha$\,Aurigae, $V=0.07\ \textrm{mag}$) 
binary has been studied for over a
century and has been observed at practically all wavelengths from X-rays
to the radio. The literature is too numerous to be cited here, but
Torres et al. (\cite{torr}) contains references
to almost all of the significant work, in particular those relevant
to the present paper. The studies of Barlow et al. (\cite{barlow}) and
Hummel et al. (\cite{hummel}) of all aspects of the astrometric
and spectroscopic orbits are cited here to emphasize their
historical importance.

Capella is one of the few binary systems with two moderately evolved
giants, most often cited as G8\,III for the primary and G0--1\,III for
the secondary. Atypical of RS~CVn binaries,  the
hotter component of Capella appears to be the more active star
rather than the cooler, usually rapidly rotating component.
This is unsurprising for its spectral type and evolutionary status as this
places the hot component in the Hertzsprung gap where rapid
redistribution of the internal angular momentum is taking place,
eventually followed by dredge up of high-angular momentum material
even to the surface. Our overall goal is to learn more about this
component by mapping its surface spots. For comparison, we previously
investigated a very similar but single G0\,III star (\object{31 Com};
Strassmeier et al. \cite{31com}). Quite unexpectedly, we found a
large polar spot in Doppler images despite the predicted
depth of its convective envelope of less than 20\%\ being much too
small to account for flux-tube deflection towards the rotational
poles. The star's rotation period was found to be 6.80$\pm$0.06 days
from MOST data. Radial-velocity variations with a full amplitude of
270~\ms\ and a period of 6.76~d were detected from our STELLA
spectra, which we also interpreted as being caused by the stellar rotation and
made us confident in the use of STELLA for this
high-precision work. The stellar mass and age for 31~Com were
determined to be 2.6$\pm$0.1~$M_\odot$ and $\approx$540~Myr from a
comparison with evolutionary tracks. To first order, we would expect
similar values for the hot secondary component of Capella.

In this paper, we report on our ongoing monitoring of Capella and
present a new improved orbital solution. The secondary's comparably
featureless spectrum with respect to the cooler component and its
large rotational line broadening of 35~\kms\ make its observation
difficult in spite of the favorable light ratio cool/hot component
of $\approx$0.9 at visual wavelengths. Our own earlier work on
Capella was based on either snap-shot spectroscopy (Strassmeier \&
Fekel \cite{str:fek}) or \Halpha\ photometry (Strassmeier et al.
\cite{str:ree}). The latter paper presented the detection of a
106$\pm$3 d photometric period for the late-G giant, which we
interpreted to be its rotation period synchronized to the orbital
motion. A period of 8.64$\pm$0.09~d was found for the G0--1 giant,
although it is less well defined and certainly less accurate than the formal error suggests
but in agreement with an independent determination from the He\,{\sc
i} equivalent widths by Katsova \& Scherbakov (\cite{kat:sch}).
However, the knowledge of accurate absolute dimensions, in
particular the component masses, radii, and rotational periods,
are essential for comparing a system's angular momentum evolution
with model predictions. Its present location in the Hertzsprung gap
suggests that there are rapid changes in its internal structure with a deepening
convection zone and associated changes in the total stellar
moment of inertia (e.g. Kim \& Barnes \cite{kim:bar}). Deep mixing
processes at this stage are also predicted to have a direct impact
on the visible surface rotation and chemical abundance
(B\"ohm-Vitense \cite{boehm}). Precise astrophysical parameters of
stars in this evolutionary stage may thus help us to understand
the angular-momentum loss in late-type stars.

We employed our new robotic spectroscopic facility STELLA in
Tenerife, Canary Islands to continuously monitor Capella for
a total of nearly four consecutive years with a cadence of one
spectrum for every useable night. The instrument and the data are
described in Sect.~\ref{S2}. A new SB2 orbit with, so far,
unprecedented accuracy is presented in Sect.~\ref{S3}. We discuss
its details in Sect.~\ref{S4}. In a forthcoming paper, we present
the first attempt to derive a Doppler image of the surface of the
rapidly rotating secondary component.

\section{STELLA/SES spectroscopy in 2007--2010}\label{S2}

High-resolution time-series spectroscopy was obtained with the
\emph{STELLA \'Echelle Spectrograph} (SES) at the robotic 1.2-m
STELLA-I telescope in Tenerife, Spain (Strassmeier et
al.~\cite{malaga}). A total of 438 \'echelle spectra were acquired
over the course of 3~1/2 years (1312~d).  The SES is a fiber-fed
white-pupil \'echelle spectrograph with a fixed wavelength format of
388--882\,nm. Despite increasing inter-order gaps in the red, it
records the range 390--720nm continuously. Its two-pixel resolution
is $R=55\,000$. The CCD is currently an E2V\,42-40 2048$\times$2048
13.5$\mu$m-pixel device.

Integrations on Capella were set to exposure times of 15~sec and
60~sec (until end of 2007) and achieved signal-to-noise ratios (S/N) 
in the range 200--500:1 per resolution element, depending on
weather conditions. The spectra were obtained between July 2007
(JD\,2,454,305.7) and March 2011. Just 14 primary and 7
secondary measurements were discarded because their data were of too low S/N due
to clouds. Our data are automatically reduced and extracted with
the IRAF-based STELLA data-reduction pipeline (see Weber et al.
\cite{spie}). The two-dimensional (2d) images were corrected for bad pixels and cosmic
rays. Bias levels were removed by subtracting the average overscan
from each image followed by the subtraction of the mean of the
(already overscan-subtracted) master bias frame. The target spectra
were flattened by dividing by the master flat, which was
normalized to unity. The robot's time series also includes nightly
and daily Th-Ar comparison-lamp exposures for wavelength calibration
and spectrograph focus monitoring. A continuous monitoring of the
environmental parameters inside and outside the spectrograph
room, most notably of temperature and barometric pressure, allows us to
apply proper corrections. For details of the \'echelle data
reduction with particular emphasis on the temperature and pressure
dependences of the SES, we refer to Weber et al. (\cite{spie}).

Twenty-two radial velocity standard stars were observed with the
same set-up and analyzed in Strassmeier et al (\cite{hd1},
\cite{orbits}). The STELLA system appears to have a zero-point
offset with respect to CORAVEL (Udry et al. \cite{udry1999}) 
of $0.50\pm0.21$~\kms, and a comparable offset  of $0.46\pm0.64$~\kms\ 
with respect to the 17 stars in common with Scarfe (\cite{scarfe2}). 
The highest external rms radial-velocity precision over the four years of observation was
around 30~\ms\ for late-type stars with narrow spectral lines. All
velocities in this paper are on the STELLA zero-point scale. The
individual velocities are listed in Table~\ref{T1}, available only
in electronic form via CDS Strasbourg.

\begin{figure}
\includegraphics[angle=0,width=\columnwidth]{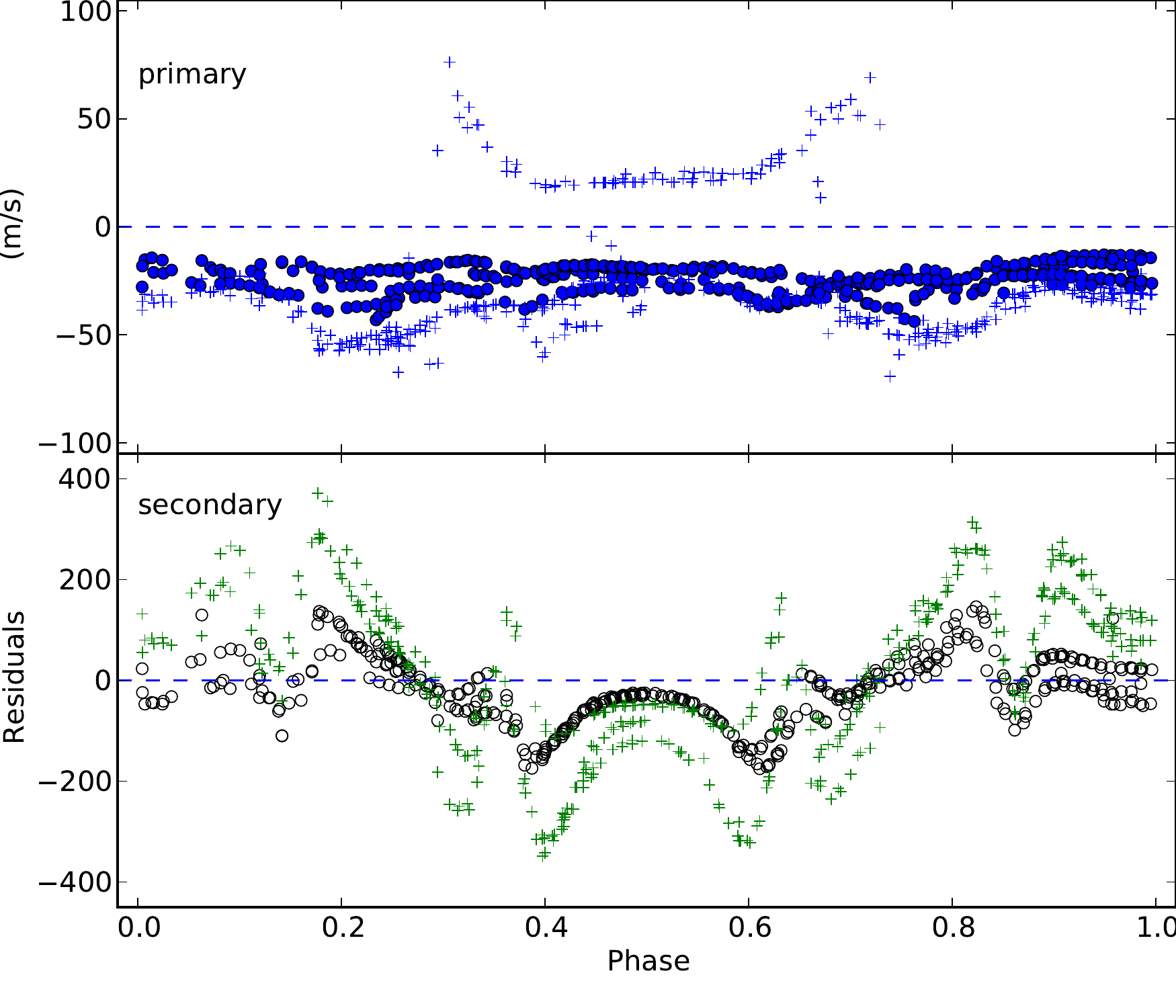}
\caption[ ]{Residual velocities from simulated spectra. Circles
denote the final setup using 62 \'echelle orders. Pluses indicate
the residuals when using only 19 orders. A simulated spectrum was
computed for the time of every observed spectrum, and the same
method as used for the observations was applied to derive the radial
velocities of the two components. The offsets shown here are the
differences between the velocities used to construct the simulated
spectra and the values measured. Systematic offsets mostly affect
the secondary star, with absolute values peaking at about 50\% of
the single measurement error of typically 300~\ms\ for component~B.
Increasing the number of orders and thus the wavelength range
clearly suppresses the systematic errors.
 \label{F2}}
\end{figure}

\begin{figure*}
\includegraphics[angle=0,width=\textwidth]{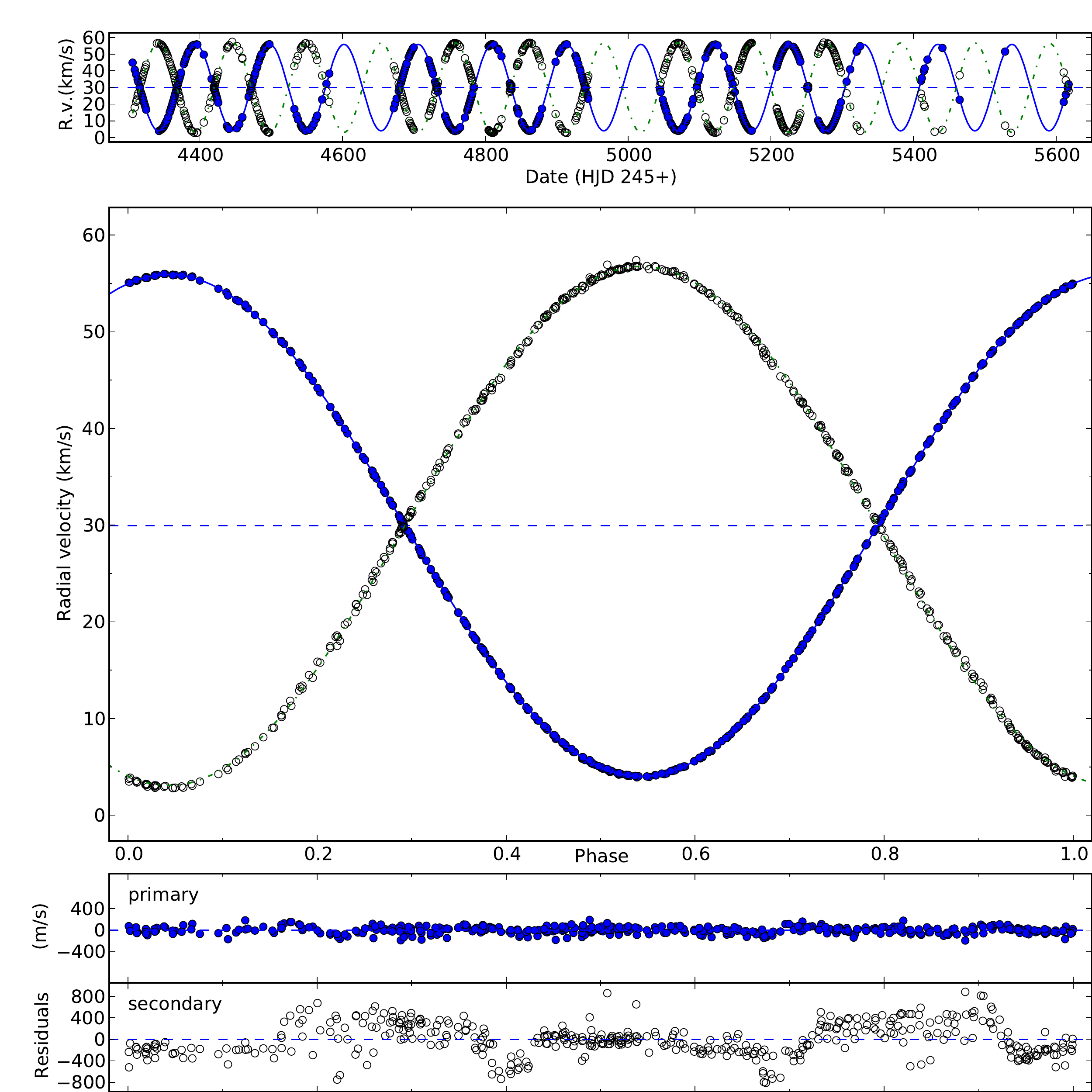}
\caption[ ]{The STELLA radial velocities compared with our newly
computed orbit.  Filled circles denote the cool component, star~A,
open circles the hot component, star~B. The two top panels show the
radial velocities versus HJD and versus orbital phase, the two
bottom panels show the residuals versus phase for the primary and
the secondary, respectively. The lines are computed from our
elements in Table~\ref{T2}. The horizontal dashed line in the middle
panel is the systemic velocity. The residuals show a phase-dependent
behavior in both components, but the overall rms of 64~\ms\ and
297~\ms\ for the two components fit the uncertainties of the
individual measurements quite well. \label{F1}}
\end{figure*}

\begin{table}
\caption{Barycentric STELLA radial velocities of Capella.}
\label{T1}
\centering
 \begin{tabular}{lllll}
  \hline \hline
HJD         & $v_{\rm A}$ & $\sigma_{\rm A}$ & $v_{\rm B}$ & $\sigma_{\rm B}$ \\
            & \multicolumn{4}{c}{(\kms)}  \\
\hline
2454305.724 	&   44.932 &	 0.029 &	   14.209 &	 0.381 \\
2454308.712 	&   40.642 &	 0.038 &	   18.032 &	 0.455 \\
\dots & \dots & \dots & \dots & \dots \\
2455616.490 	&   30.384 &	 0.034 &	   29.515 &	 0.222 \\
2455617.482 	&   31.949 &	 0.030 &	   28.070 &	 0.254 \\
\hline
\end{tabular}
\end{table}

\section{New spectroscopic orbital elements}\label{S3}

\subsection{High-precision radial velocities}
\label{orbit}

During the initial year of STELLA operations, the external rms values
were significantly larger (120~\ms ) than thereafter (30~\ms
) and included a radial velocity offset of about 300~\ms . The final
radial velocities are barycentric and corrected for Earth's
rotation.

The STELLA velocities listed here are determined from a simultaneous
cross-correlation of 62 \'echelle orders with a proper synthetic
spectrum. To perform this, we selected from a pre-computed grid of synthetic
spectra from ATLAS-9 atmospheres (Kurucz~\cite{kur}) two spectra
that match the respective target spectral classifications. For
Capella A, we chose the $T_{\rm eff}=5000\ \mathrm{K}$ model, and for Capella
B the $T_{\rm eff}=5500\ \mathrm{K}$ model, both with $\log g=3.0$ and solar
metallicity. We combined the two template spectra to one artificial
spectrum using $v\sin i$ of 4.5~km\,s$^{-1}$ and 35~km\,s$^{-1}$ for
the A and B component, respectively, and a macroturbulence of
5~km\,s$^{-1}$. We adjusted the (wavelength dependent) intensity
ratio using G8\,III and G0\,III flux-calibrated spectra from
Pickels (\cite{pickels}) multiplied with 0.955. We then applied a
series of radial velocity differences between the two components,
and computed a two-dimensional cross-correlation function for each
of these shifts. The highest correlation value in the resulting
two-dimensional image corresponds to the measured velocity of
component A in one dimension, and the velocity difference of the two
components in the other dimension. To estimate the uncertainties in
this method, we evaluated a series of 1000 Monte-Carlo variations in
these two-dimensional images using only 63\% of the original number
of orders, estimating $\sigma$ to be 1/1.349 of the interquartile
range of the resulting distribution.

To correct for systematic radial velocity offsets induced by the
measuring technique, we created simulated stellar spectra from the
above templates at exactly the same wavelength spacing as the original
observations and corrected the measured radial velocities with these
offsets. We initially performed our analysis with 19 \'echelle
orders but the final analysis was performed with all usable orders, which
turned out to be 62 out of 82 in our case. Fig.~\ref{F2} illustrates
the systematic errors and their dependance on the number of orders,
19 compared to 62, and the wavelength range used. Doing the analysis with
just 19 orders resulted in a $K_\textrm{B}$ of about 400~\ms\
smaller than with 62 orders, even when taking the corrections from
Fig.~\ref{F2} into account. This shows that the systematic effects
are not fully taken into account by the simulations.

We also correct the radial velocities of the B component for the
difference in gravitational redshift according to Lindgren \&
Dravins (\cite{lindgren:dravins}). For component~A, this amounts to
132~\ms\ and for component~B to 177~\ms\ with the masses and radii
from Table~\ref{T2}, i.e. a net differential offset of 45~\ms.
After applying this correction, a residual offset of the respective
center-of-mass velocities of the two components of 59~{\ms}
remained. This small offset could be caused by the different
line-shapes of the two stars, e.g. produced by either convective blue- or
redshifts (see e.g. Gray \cite{gray2005}). Therefore, we applied a
total correction of 104~\ms\ to the radial velocities of the B component. The main
effect of this correction is a small decrease in the secondary's
velocity amplitude. We note that all of above corrections still amount
to changes in the final parameters of smaller than 1$\sigma$.

\subsection{Spectroscopic orbit}

Figure~\ref{F1} presents our new STELLA velocities and compares them
with the computed velocity curves.

We solved for the usual elements of a double-lined spectroscopic
binary using the general least-squares fitting algorithm {\em MPFIT}
(Markwardt \cite{mpfit}), utilizing the reciprocal values of $\sigma$ from Table~\ref{T1} as weights. 
For solutions with non-zero eccentricity,
we used the prescription of Danby \& Burkardt
(\cite{danby:burkardt}) to calculate the eccentric
anomaly. The orbital period was fixed at 104.02173~d (Torres et
al.~\cite{torr}). We initially assumed a zero eccentricity, but
the final corrected data yields a best-fit solution at $e = 0.00087 \pm
0.00013$ and $\omega = 345 \pm 9.7$ for both the SB2 solution and
the solution of the primary's velocities alone. Using ony the
secondary's velocities, we obtain a zero eccentricity. The rms of the
eccentric solution is 64~\ms\ and 297~{\ms} for the primary and
secondary velocities, respectively, which is slightly better than
the 65~\ms\ and 301~\ms\ obtained from the circular solution. The
eccentricity is also consistent with the values of $e = 0.00083 \pm 0.00005$ and
$\omega = 334.8 \pm 4.7$ that Torres et al. (\cite{torr}) derived
from the astrometric measurements of Hummel et al. (\cite{hummel}),
and practically identical to their result when all radial-velocity
and astrometric data were combined ($0.00087 \pm 0.00021$ and $324
\pm 14$).

\begin{table}[tbh]
\centering
\caption{Spectroscopic orbital elements of Capella.}\label{T2}
\begin{tabular}{lll}
\hline \hline
Parameter & Our Value & Torres et al.\\
\hline
$P$ (days)                      & &104.02173$\pm$0.00022\\
$T_{\rm Periastron}$ (HJD 24+)      & 54\,389$\pm$2.8 & 47\,528.514$\pm$0.016\\
$\gamma$ (km~s$^{-1}$)          & 29.9378$\pm$0.0026 &  29.653$\pm$0.035\\
$K_A$ (km~s$^{-1}$)             & 25.960$\pm$0.0065 & 26.005$\pm$0.036\\
$K_B$ (km~s$^{-1}$)             & 26.840$\pm$0.024 & 26.260$\pm$0.087 \\
$e$                             & 0.00087$\pm$0.00013 & 0.0\\
$\omega$ (deg)                  & 345$\pm$9.7 & - \\
$a_A$~sin~$i$ (10$^6$ km)       & 37.134$\pm$0.0093  & \\
$a_B$~sin~$i$ (10$^6$ km)       & 38.392$\pm$0.034 &\\
$M_A$~sin$^3$~$i$ ($M_{\odot}$) & 0.8064$\pm$0.0014 & \\
$M_B$~sin$^3$~$i$ ($M_{\odot}$) & 0.7800$\pm$0.0008 & \\
$N_A$, $N_B$\tablefootmark{a}             & 424, 431 & 504, 162\tablefootmark{b} \\
rms$_A$ (m\,s$^{-1}$)           & 64 &\\
rms$_B$ (m\,s$^{-1}$)           & 297 &\\
mass ratio, $q\equiv M_B/M_A$   & 0.9673$\pm$0.0020 & 0.9903$\pm$0.0036\\
inclination, $i$ (deg)         & & 137.21$\pm$0.05 \\
$M_A$ ($M_\odot$)               & 2.573$\pm$0.009  & 2.466$\pm$0.018 \\
$M_B$ ($M_\odot$)               & 2.488$\pm$0.008 & 2.443$\pm$0.013 \\
$R_A$ ($R_\odot$)            &   & 11.87 \\
$R_B$ ($R_\odot$)            &   & 8.75 \\
\hline
\end{tabular}
\tablefoot{\tablefoottext{a}{$N$ is the number of measurements for each component.}
\tablefoottext{b}{A total number of 1015 measurements were used including astrometry.}}
\end{table}

\section{Discussion and conclusions}\label{S4}

Most prior studies of the spectroscopic orbit of Capella have found
that the early G star is the least massive component in the Capella
system. The only exception is an early orbit by Struve \& Kung
(\cite{str:kun}) from McDonald-Observatory spectra, as noted by
Barlow et al. (\cite{barlow}). Torres et al. (\cite{torr}) collected
most previously obtained spectroscopic orbits and documented the
decrease in the measured semi-amplitude of the secondary with ever
increased observational precision. The difficulty in measuring the
radial velocity of the secondary (of a zero magnitude star!) is
thereby nicely demonstrated. The earliest orbital solutions had a
secondary to primary mass ratio of $\approx$0.74, which became
$\approx$0.87 by the mid eighties in the past century, $\approx$0.94
in the nineties, and 0.9903$\pm$0.0036 in the study of Torres
et al. (\cite{torr}). In the present paper, we verify the almost
unity mass ratio but confine it to 0.9673$\pm$0.0020 and derive
masses good to 0.3\%.

The two previously highest quality data sets -- Torres et al. (\cite{torr}) from
CfA observations, and Barlow et al. (\cite{barlow}) from a
combination of DAO, McDonald, and KPNO observations -- have
comparable rms for the primary of 0.46 and 0.47~\kms\ but 0.91 and
1.38~\kms\ for the secondary, respectively. The CfA data set appears
to be significantly more precise for the secondary velocities than
the combined Barlow et al. (\cite{barlow}) velocities. However, the
CfA velocities are not free of systematics. Firstly, they were
obtained from single-order \'echelle spectra with a wavelength
coverage of just 45~\AA , 117 times smaller than our STELLA/SES data.
As shown by the authors themselves, this introduces phase-dependent
radial velocity variations of the secondary star of up to
$\pm$1~\kms, which are produced by residual blending and spectral lines moving in
and out of the cross-correlation window. All CfA velocities had to
be corrected for this effect before further use. Secondly, when the
CfA velocities from both components were solved for separately,
their best solution was found for $\gamma$-velocities that differ by
0.27$\pm$0.08~\kms , with the B-component value being the smaller one. This
is not uncommon in double-lined solutions but this difference also
persisted in their global solution. Spot activity on both
stars can account for such an effect but should statistically cancel
out for a long enough database timeline unless phenomena such as active spot
longitudes or variable spot lifetimes play a role.

Our STELLA data are not affected as strongly by the above
problems, the precision of an individual measurement even when
compared to the best previous data being already higher by a factor
3--4, owing to spectrograph stability and higher spectral
resolution. Our rms from the orbital solution is 64~\ms\ for the
primary and 297~\ms\ for the secondary. The sampling is also
unprecedented and a total of 424 velocities for the primary and 431
for the secondary were available for our analysis. If we use the
best-fit orbital period from the STELLA data alone, i.e.
104.0243$\pm$0.0007~d, we find a practically indistinguishable orbit
from that listed in Table~\ref{T2} with rms-values of within 1~\ms and 4~\ms for
the cool and hot components, respectively. Because our time coverage
is shorter than that considered in the combined Torres et al.\
orbit, 1312~d versus 5037~d, we chose to adopt their orbital period
from their combined solution. Taking our velocity offset of
+0.503~\kms\ with respect to CORAVEL into account, our $\gamma$
velocity is smaller by 0.218~\kms\ than the Torres et al. value of
29.653$\pm$0.053~\kms\ but, more importantly, more precise by a
factor~20. The semi-amplitude of component~A from STELLA is smaller
by just 0.045~\kms\ but more precise by a factor of 5.5, while for
component~B it is larger by 0.580~\kms\ and more precise by a factor
of 3.6. As noted in Sect.~\ref{orbit}, we obtain a $K_{\rm B}$ value
within the error bars similar to the one found by Torres et al.\  when using
only 19 spectral orders. We therefore conclude that the systematic
errors in the secondary velocity make $K_{\rm B}$ appear smaller
(see Fig.~\ref{F2}). Because of the above rms numbers and that
an inspection of the radial-velocity curves in Fig.~\ref{F1} shows
residual systematic errors only at the level of the accuracy of a
single measurement, we conclude that our orbit is also more
accurate. We speculate that the inclusion of the many different data
sets in the Torres et al. solution, spectroscopic as well as
astrometric, and their simultaneous global approach with a total of
26 free adjustable parameters creates high precision but lesser than
the hoped for accuracy, possibly because of remaining systematics in the
individual data sets.

The mass ratio of a binary with two evolved giant components is of
critical importance for the determination of their evolutionary
status. This is particularly obvious for the Capella system as we
have the case where two practically equally massive giants occupy
rather different locations in the H-R diagram, and yet have the
same age. Mostly by virtue of its rotational line broadening, the hotter B
component is the one with the more uncertain parameters. Its
spectral classification, however, unambiguously places it within the
Hertzsprung gap where rapid changes in terms of surface chemistry,
angular momentum etc. are taking place. The orbital elements from
Table~\ref{T2}, combined with the inclination of the orbital plane
from the astrometric orbit of 137.21\degr\ (again cf. Torres et
al.), give masses of 2.573$\pm$0.009~$M_\odot$ and
2.488$\pm$0.008~$M_\odot$ for the A and B components, respectively.
These are masses with relative errors of 0.34\% and 0.30\% , about
half of the errors given in Torres et al., and with absolute values
that differ by 4\% and 2\% for the two stars. We note that we continue to
denote the A component as the primary because it is the dominant
source in the combined spectrum but also recall that the secondary
is the brighter of the two components at optical wavelengths
(Griffin \& Griffin \cite{g:g}, Strassmeier \& Fekel
\cite{str:fek}).

\acknowledgements{STELLA was made possible by funding through the
State of Brandenburg (MWFK) and the German Federal Ministry of
Education and Research (BMBF). The facility is a collaboration of
the AIP in Brandenburg with the IAC in Tenerife. We thank all
engineers and technicians involved, in particular Manfred Woche
and Emil Popow and his team as well as Ignacio del Rosario and
Miquel Serra from the IAC Tenerife day-time crew.
We thank the referee, Frank Fekel, for his constructive critisism. 
}

\end{document}